\documentclass[twocolumn,preprintnumbers,amsmath,amssymb,A4paper]{revtex4}

\usepackage{graphicx}
\usepackage{dcolumn}
\usepackage{bm}

\begin{document}

\newcommand{\kp}{{\bf k$\cdot$p}\ }
	
	\title{Theoretical model of a new type tunneling transistor}
	\author{Pawel Pfeffer\footnote{pfeff@ifpan.edu.pl}}

\address{Institute of Physics, Polish Academy of Sciences, Aleja Lotnikow 32/46, PL-02668 Warsaw, Poland}

\begin{abstract}
	A tunneling transistor without heterojunction as a theoretical design, or more precisely controlled electron current transmission by barrier potential, is under consideration.
	The electrons from the conduction band of the source tunnel through the forbidden gap $E_g$ of the channel to the conduction band of the drain. The tunneling current $J$ calculations made at helium temperature for the example InAs-InAs-InAs, Au-GaSe-Au and Al-AlN-Al structures show that for a constant source-drain voltage, $V_C$, of several mV, changes in the gate voltage, $V_G$, applied to the channel within the voltage range of 0 - $E_g/$2e change $J$ by even 10 orders of magnitude. Unlike the existing solutions such as tunnel field-effect-transistor (TFET), the proposed device uses the change of $V_G$ (gate voltage), i.e. the change of the electrostatic potential in the channel, to modify the imaginary wave vector $k_z$ of tunnel current electrons. Consequently, the gate voltage controls the damping force of the electrons wave functions and thus the magnitude of the tunneling current, $J$. The effect of increasing temperature, T, on $J(V_G)$ relation was also tested. It was found that only in structures with a wide forbidden channel gap  this effect is insignificant (at least up to T=300 K).
	
\end{abstract}
	\maketitle

\section{\label{sec:level1} Introduction\protect\\ \lowercase{}}

\begin{figure}
\includegraphics[scale=0.4,angle=0, bb = 750 80 40 550]{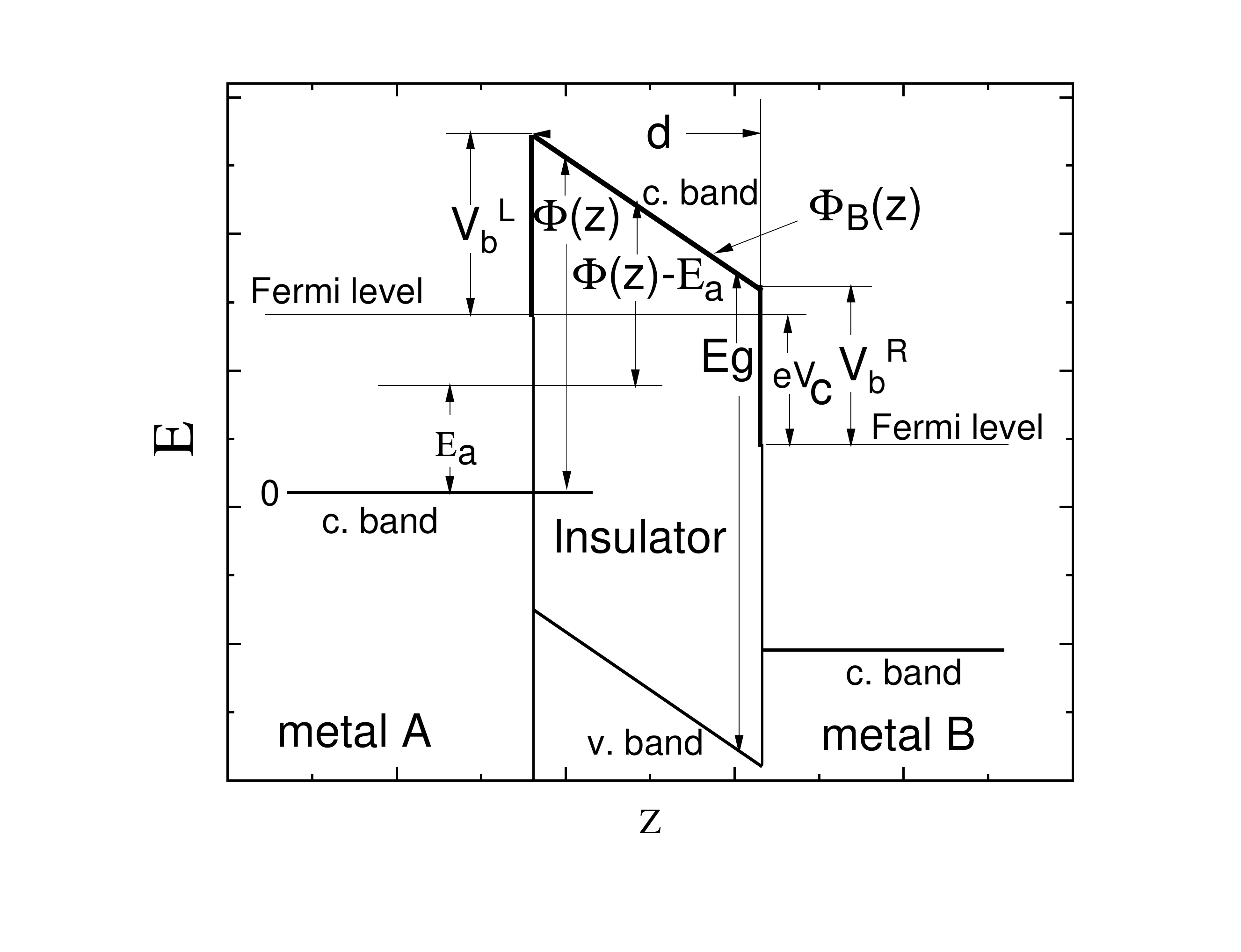}	
\caption{Diagram of the Metal-Insulator-Metal structure with applied voltage $V_C$ between the metal A and the metal B, see e. g. Ref.\cite {St}. The tunneling current flows along the $z$ direction from the metal A through the forbidden gap of the insulator to the metal B. $V_b^L$ and $V_b^R$ are the band offsets between the conduction band of the insulator and the Fermi energy $E_F$ in the metal A or the metal B, respectively. $E_a$ is the energy of the electron.} \label{fig1th}
\end{figure}

The tunnel current in the Metal-Insulator-Metal (M-I-M) structure has been theoretically described and experimentally tested for many years, starting in 1930 Ref.\cite {Fr1930}, see also e.g. Refs.\cite {Si}-\cite {Zh}. The theoretical problem of the transmission of relativistic electrons through the potential barrier of a controlled height, described by the Dirac equation, called the Klein paradox, was published Ref.\cite {Kl} and resolved many years ago, see review  Ref.\cite {Ca}. The similar theoretical problem, in close analogy to Dirac model, but for transmission coefficient of conduction electrons through the potential barrier with the height depending on the voltage applied to it, $V_G$ (gate potential), was solved for graphene Ref.\cite{Ka}, for phosphorus Ref.\cite{Li2017} and for IV-VI semiconductor compounds Ref.\cite{Pf} in recent years. Both of these phenomena are the basis of the proposed theoretical tunneling transistor model. Thereby, the considered design, described below, differs from existing TFET solutions based on the fact that an increase in $V_G$ means an increase of
the bands bending in the source-channel heterojunction and, as a result, an increase in the tunnel current $J$ from the valence band to the conduction band in $p^+-i-n^+$ or $p^{++}-n^--n^+$ configuration or from the conduction band to the valence band in $n^+-i-p^+$ or $n^{++}-p^-p^+$ configuration (band-to-band tunneling (BTBT)) , see e.g. Refs.\cite{Sa}-\cite{Co}.
\section{\label{sec:level1} Tunneling Current in Metal-Insulator-Metal structure\protect\\ \lowercase{}}

Here we consider details of the general formula for the dependence of the current $J$ on the applied voltage, $V_C$, in the M-I-M structure (see Ref. \cite{St}) as the source-channel-drain structure in our design. The elements of this structure are selected so that current electrons from the metal A tunnel along the $z$ direction through the forbidden gap of the insulator to the metal B, see Fig. 1. The wave vector $k_z(z)$ of the tunneling electrons has a decisive influence on the magnitude of the $J$ current. This vector has an imaginary value and determines how much the electron wave function is damped in the insulator area.
\begin{figure}
	\includegraphics[scale=0.35,angle=0, bb = 650 5 50 500]{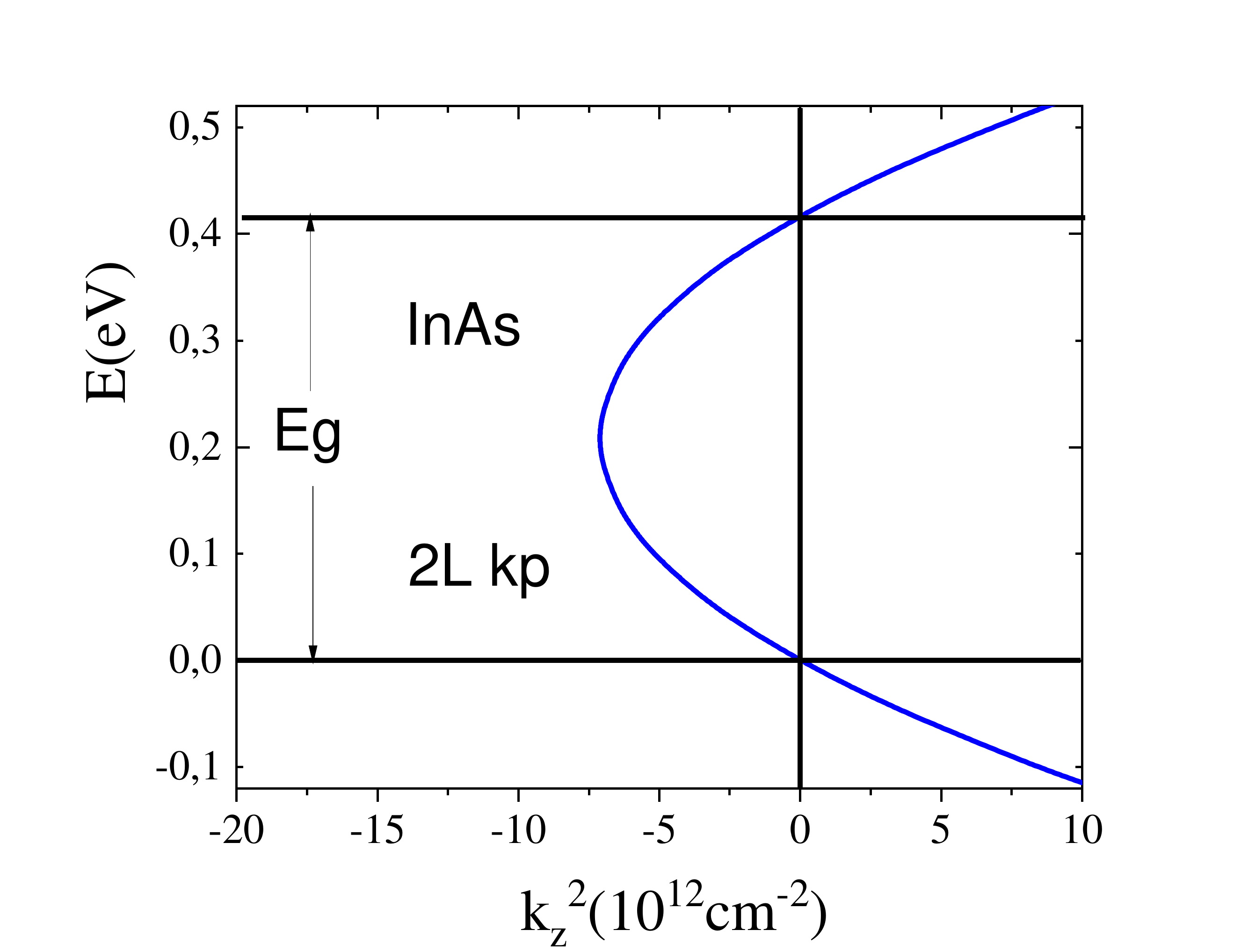}
	\caption{The theoretical energy dispersion $E(k_z^2)$ in the forbidden gap of InAs for $k_{\perp}$ = 0 calculated in the two level model for $E_g$ = 0.417 eV and $m^*_0/ m_0 = 0.026$.} \label{fig2th}
\end{figure}

In order to determine the formula for $|k_z(z)|$ we proceed as follows:
the energy of electron in the insulator, $E(k)$, counted from the bottom of the forbidden gap, in the two band model, is:
\ \\
$$
E(k) = E_g-((\Phi(z)-E_a) =
$$
\begin{equation}
	=\frac{E_g}{2} \pm
\left[(\frac{E_g}{2})^2+\frac{E_g\hbar^2({k^i_z}^2+k^2_{\perp})}{2m^*_0}\right]^{1/2}\;\;,
\end{equation}
where $E_g$ is the forbidden gap of the insulator, $k^2= k^2_{\perp} + (k_z^i)^2$, and wave vectors $k_{\perp}$ and $k_z^i$ have real and imaginary value in $E_g$, respectively. $E_a$ is the energy of the electron in the metal A, $m^*_0$ is the effective electron mass $m_c$ at the conduction-band edge or $m_v$ at the valence-band edge, corresponding to the sign in front of the square root in Eq.(1) and $V_C$ is the applied voltage. $\Phi(z) = \Phi_B(z)+E_F$ is the energy of the M-I-M barrier potential relating to the metal A conduction band edge, where
\begin{equation}
\Phi_B(z) = V^L_b+(V^R_b-V^L_b-e V_c)z/d\;\;,
\end{equation}
$ V^L_b$ and $ V^R_b$ are metal-insulator barrier energies and $E_F$ is the Fermi energy.

Hence,
\begin{equation}
|k_z(z)| = \left[\left(1-\frac{\Phi(z)-E_a}{E_g}\right)(\Phi(z)-E_a)\frac{2m^*_0}{\hbar^2} + k^2_{\perp}\right]^{1/2}\;\;.
\end{equation}
or
\begin{equation}
|k_z(z)| = \left[\left(1-\frac{E(z)}{E_g}\right)E(z)\frac{2m^*_0}{\hbar^2} + k^2_{\perp}\right]^{1/2}\;\;.
\end{equation}
\begin{figure}
	\includegraphics[scale=0.35,angle=0, bb = 650 5 90 520]{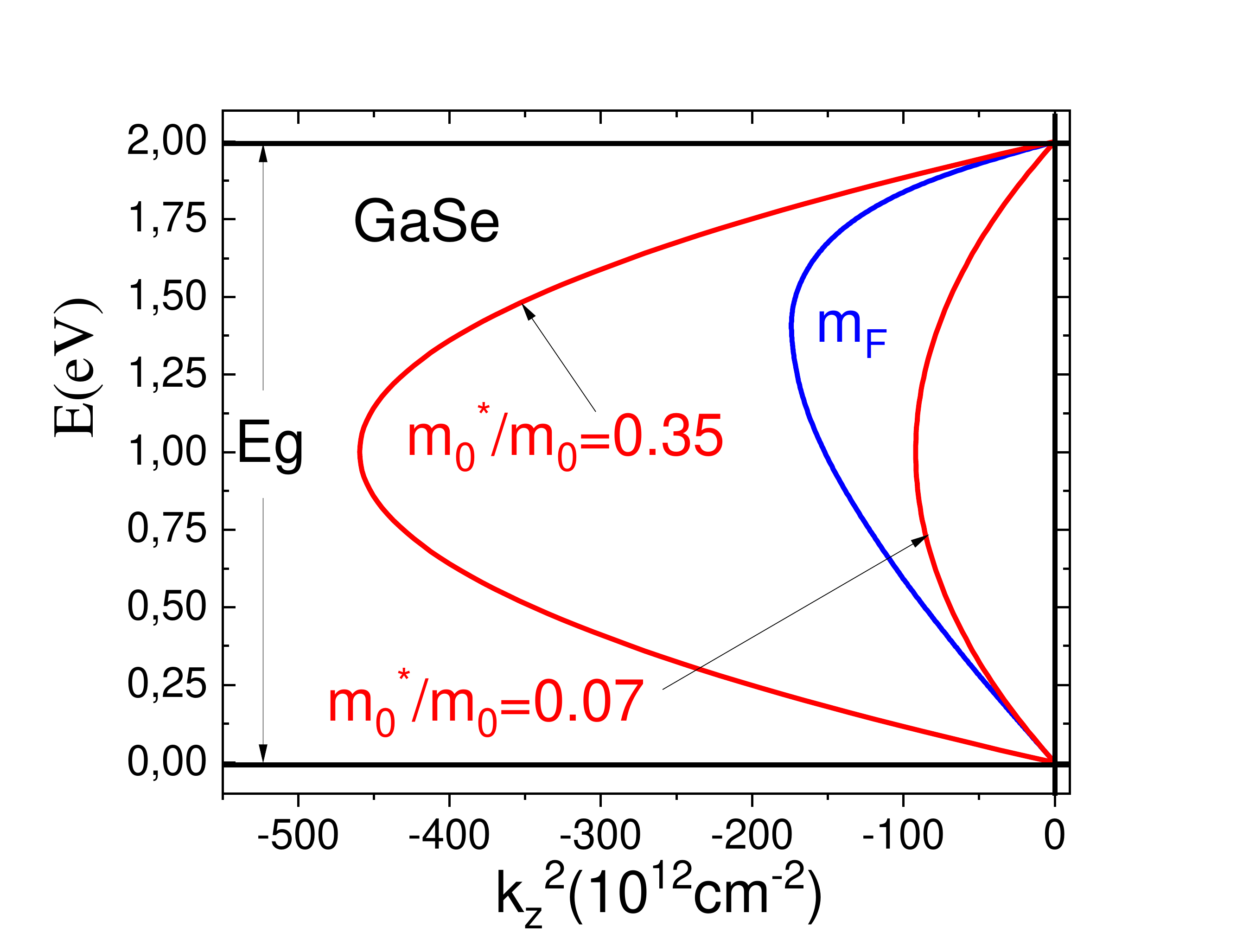}
	\caption{The theoretical energy dispersions $E(k_z^2)$ in the forbidden gap of GaSe for $k_{\perp}$ = 0. The blue curve includes the electron mass $m_F(E)$ calculated in the two level Franz model for $m_C/m_0=0.35$ and $m_V/m_0=0.07$ $E_g$ = 2 eV. The red curves are calculated in the two level model for $m^*_0/m_0=0.35$ or $m^*_0/m_0=0.07$.}	
	\label{fig3th}
\end{figure}

The next step is a general expression for the elastic tunneling current J(V) from the metal A to the metal B (see e. g. Ref \cite{Ku}),
$$
J(V) = \frac{2eS}{\hbar}\int_0^{\infty}dE_a(f_A(E_a)-f_B(E_a))\cdot
$$
\begin{equation}
\cdot\int_0^{\infty}\frac{d^2k_{\perp}}{(2\pi)^2}\rm{exp}\left[-2\int^d_0|k_z|(k_{\perp}, \Phi(z)-E_a)dz\right]=
\end{equation}
$$
=\frac{2eS}{\hbar}\int_0^{\infty}dE_a(f_A(E_a)-f_B(E_a))\cdot
$$
$$
\cdot\int_0^{\infty}\frac{dk_{\perp}k_{\perp}}{2\pi}\rm{exp}\left[-2\int^d_0|k_z|(k_{\perp}, \Phi(z)-E_a)dz\right]\;\;,
$$
or
$$
J(V)/S\left[\frac{A}{cm^2}\right]=\frac{7.7483}{10^5}\int_0^{\infty}dE_a(f_A(E_a)-f_B(E_a))\cdot
$$
\begin{equation}	
\cdot\int_0^{k^{M}_{\perp}} dk_{\perp}k_{\perp}
\rm{exp}\left[-2\int^d_0|k_z(z)|dz\right]
\end{equation}
where $k^{M}_{\perp} = k^F_{\perp}$ for $E_F$ and $k_z$=0 in the M-I-M system. Further, $d$ is the insulator thickness and S is the area of the interface between the metal and the insulator. $f_A(E_a)$ and $f_B(E_a)$ are F-D distribution functions for the metal A and the metal B,
$f_A(E_a) = 1/[1+\rm{exp}(E_a-E^F_a)/kT]$ and $f_B(E_a) = 1/[1+\rm{exp}[E_a-(E^F_a-eV_C)]/kT]$.

For very low temperature one has
$$
J(V)[\frac{A}{cm^2}]  = \frac{7.7483}{10^5}\cdot
$$
\begin{equation}
\cdot\int_{E_F-eV}^{E_F}dE_a\int_0^{k^{M}_{\perp}} dk_{\perp}k_{\perp}	\rm{exp}\left[-2\int^d_0|k_z(z)|dz\right]\;\;,
\end{equation}

One can notice that the formula for $J(V)$ is dominated by the element with the exponential decay which is a result of the imaginary value of $k_z$ in the electron wave function for the insulator forbidden gap.
It is also seen that all electrons in the metal A with energy in the range $E_F - (E_F-eV_C)$ form the tunneling current from the metal A to the metal B. Furthermore, the tunneling current of the electron is the bigger the smaller $|k_z|$ it has.
\section{\label{sec:level1} Dependence of current electrons energy on $k^2_z$ in channel forbidden gap \protect\\ \lowercase{}}
\begin{figure}
	\includegraphics[scale=0.35,angle=0, bb = 650 10 60 550]{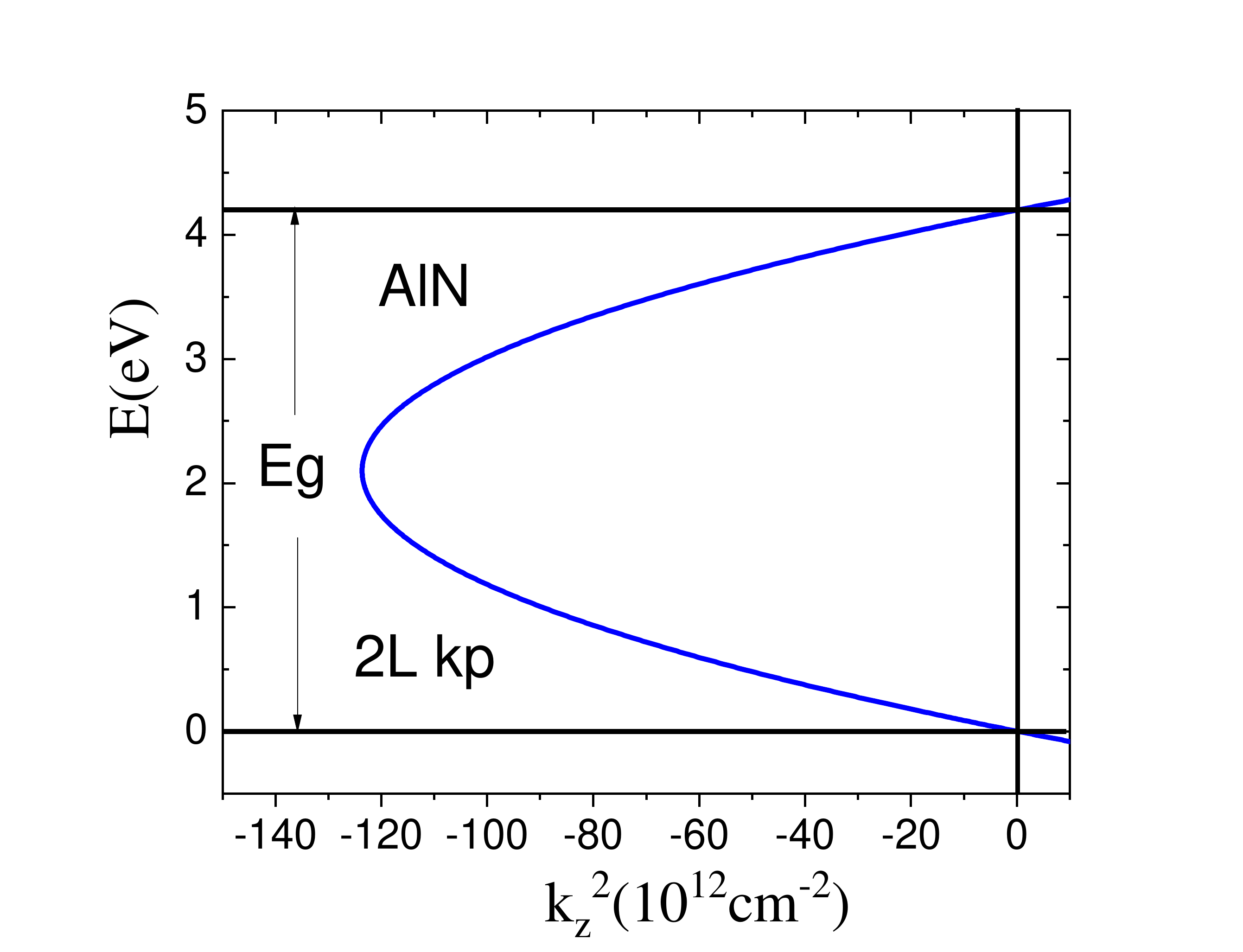}
	\caption{The theoretical energy dispersion $E(k_z^2)$ in the forbidden gap of AlN for $k_{\perp}$ = 0 calculated in the two level model for $E_g$ = 4.2 eV and $m^*_0/ m_0 = 0.45$.
	} \label{fig4th}
\end{figure}

Henceforth, the term Metal-Insulator-Metal is replaced by the term Source-Channel-Drain, source and drain are metals or $n$ type semiconductors and channel is a wide, medium or narrow gap semiconductor.

The comparison of experimental data with theoretical calculations of $J(V)$ in the structure under consideration shows that to describe the dispersion $E(k_z^2)$ of electronic states in the forbidden gap of a channel, for example InAs, GaSe or AlN, the two-band model is sufficient, see Refs. \cite{Pa}, \cite{Ku} and \cite{Gill}. The knowledge of the dependence $E$ on $k_z^2$ for $k_{\perp}$ = 0 is the most important, because the greater the $k_{\perp}$, the greater the $|k_z|$ for keeping the electron energy unchanged. On the other hand, the tunnel current is determined by the  electrons with $|k_z|$ as small as possible, i.e. with the damping of their wave function as little as possible.

If the effective masses of electrons $m_c$ and
holes $m_v$ are not equal the use of the two band Franz model for band-to-band tunneling, see Refs \cite{Fr1952} and \cite{Ta}, allows for a more detailed description of the tunneling process. It means replacing $m^*_0$ in Eq.(1) by $m_F$, the value of which depends on the energy of the electron $E$ in the band forbidden gap. $m_F$ has the form
\begin{equation}
	m_F(E) = \frac{m_c}{(E/E_g)(1-m_c/m_v)+m_c/m_v}\;\;,
\end{equation}\ \\
where $E = E_g-\Phi(z)+E_a$, see Fig. 1. It is seen that for $E = 0$ $m_F = m_v$ and for $E = E_g$ $m_F = m_c$.
\begin{figure}
	\includegraphics[scale=0.35,angle=0, bb = 650 10 60 560]{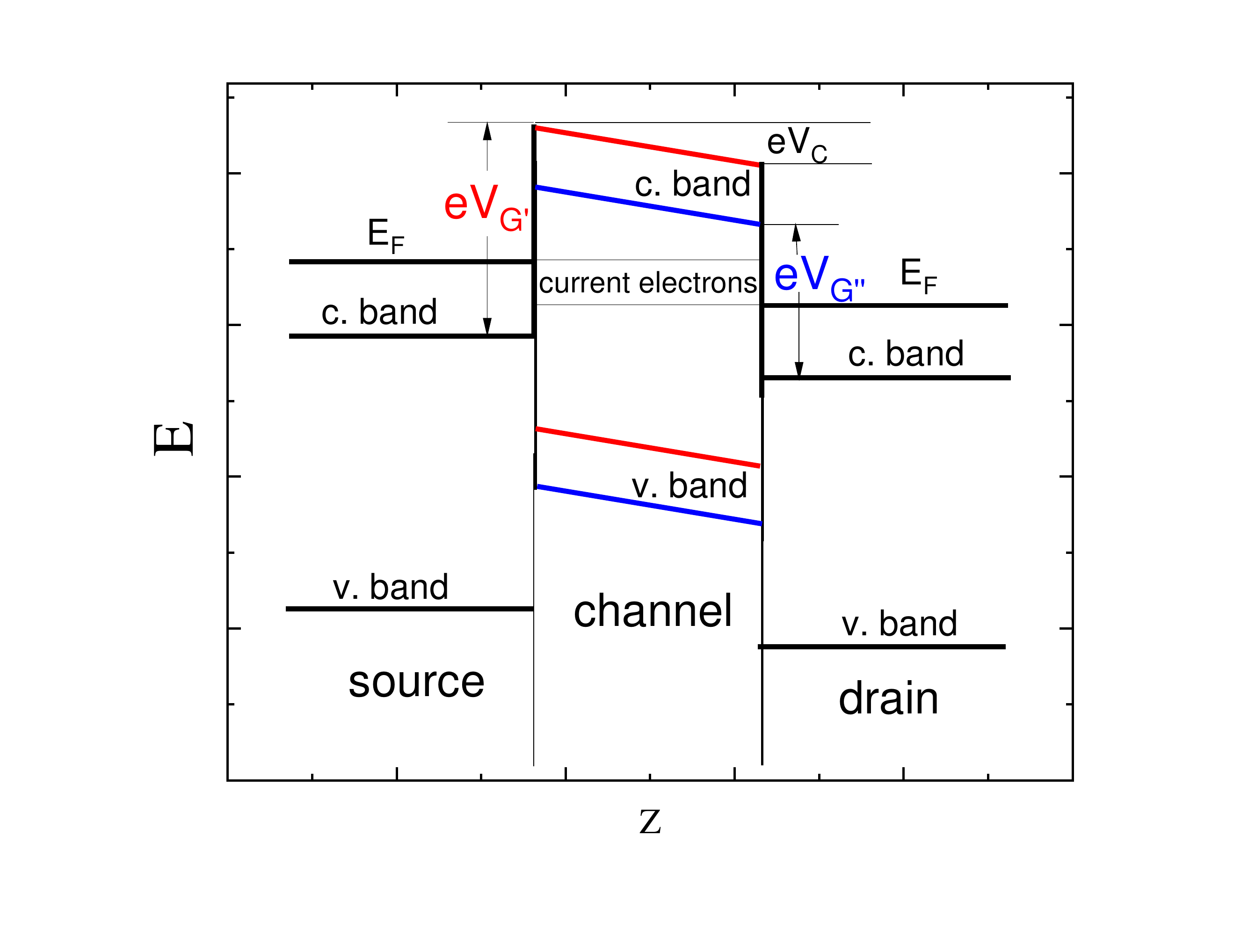}
	\caption{Scheme of the proposed tunneling transistor for the source-drain voltage $V_C$ and for two different values of the gate voltage $V_{G'}$ and $V_{G''}$ applied to the channel. It is seen that the current electrons with energy in the range $E_F$ - ($E_F$ - $eV_C$) tunnel through a barrier with a $V_G$-dependent height, i.e. through energetically different parts of the forbidden gap of the channel.} \label{fig5th}
\end{figure}

To calculate the energy dispersion $E(k_z^2)$ in the forbidden gap of GaSe for $k_{\perp}$ = 0 we used Eq. (3) with $m^*_0 = m_F$ and GaSe parameters (Ref.\cite{Ku}) $ E_g $ = 2 eV $m_C/m_0=0.35$ and $m_V/m_0=0.07$. The curves calculated for $m^*_0 = m_F$, $m^*_0 = m_C$ and $m^*_0 = m_V$ are shown in Fig. 3. From comparison of the curves it follows that the use of $m_F (E)$ is necessary. The results of similar calculations for InAs and AlN are shown in Figs. 2 and 4 respectively. InAs parameters are  $E_g $ = 0.417 eV and $m^*_0/ m_0=0.026$ while AlN parameters are $E_g $ = 4.2 eV and $m^*_0/ m_0=0.45$.
Figs. 2-4 show that a slight change in the energy of the electron in the band gap significantly changes the value of $k_z^2$ of the electron, i.e. its importance in the formation of the tunnel current.
\begin{figure}
\includegraphics[scale=0.35,angle=0, bb = 700 10 70 500]{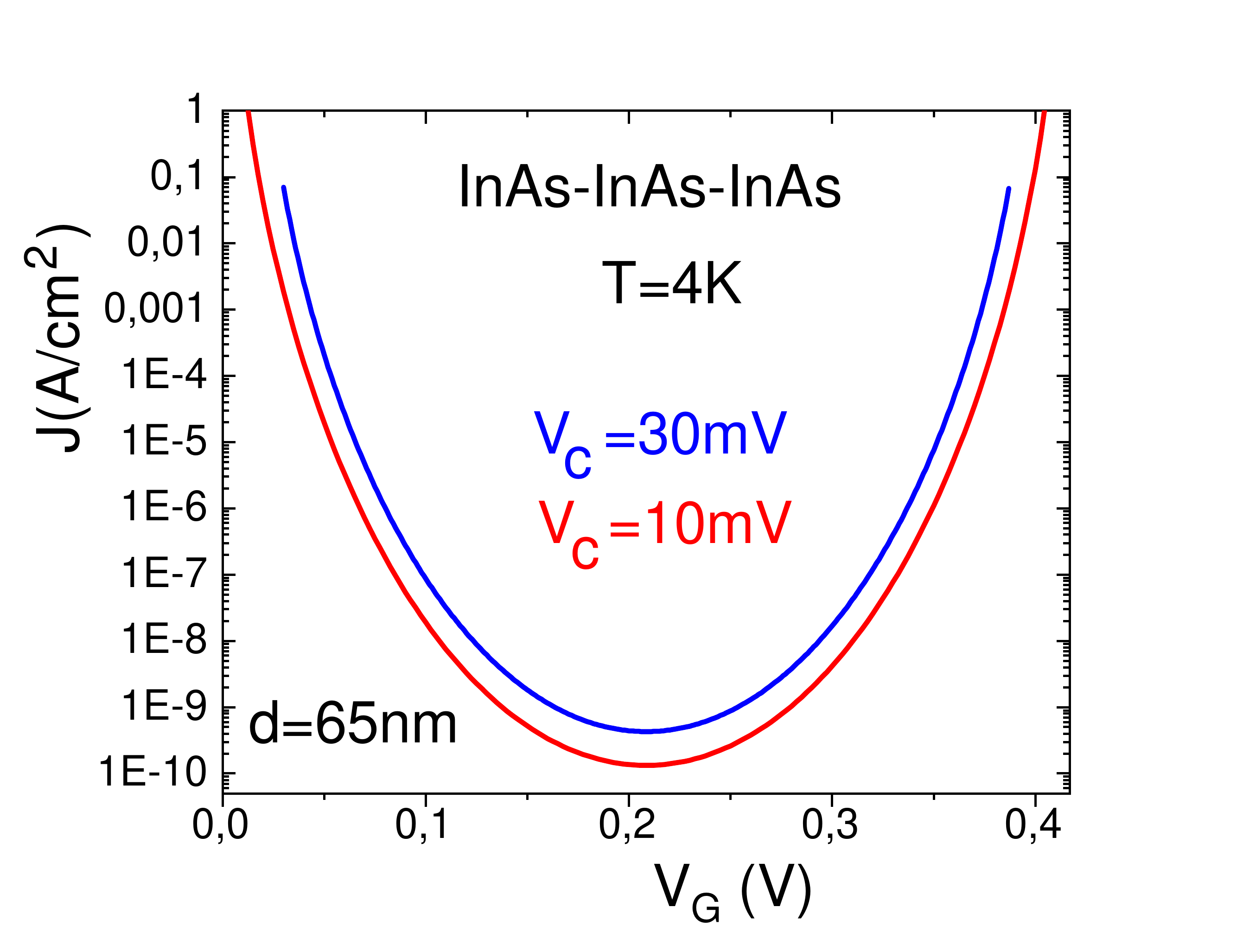}
\caption{The theoretical tunneling current $J$ versus gate voltage $V_G$ applied to the InAs barrier. The curves are calculated for values of voltage $V_C$ = 30 meV and 10 meV applied to the InAs-InAs-InAs structure with the width of the barrier $d$ = 65 nm. For $V_G$ in the range $V_C$ - (${E_g}/e- V_C$), the electrons tunnel through the whole width of the InAs forbidden gap, see Fig 1. A change of the value of $V_G$ shifts the InAs barrier on the energy axis, thereby changing the value of energy, $E(k_z)$, and the value of the wave vector, $k_z$, of tunneling electrons in the forbidden gap, (see Fig. 2) and, as a result, leads to the exponential change in the value of the tunneling current, see Eq. (6). It is seen that a slight change of $V_G$ can change the value of the tunneling current $J$ by a few orders of magnitude.} \label{fig6th}
\end{figure}
\begin{figure}
\includegraphics[scale=0.35,angle=0, bb = 650 10 60 500]{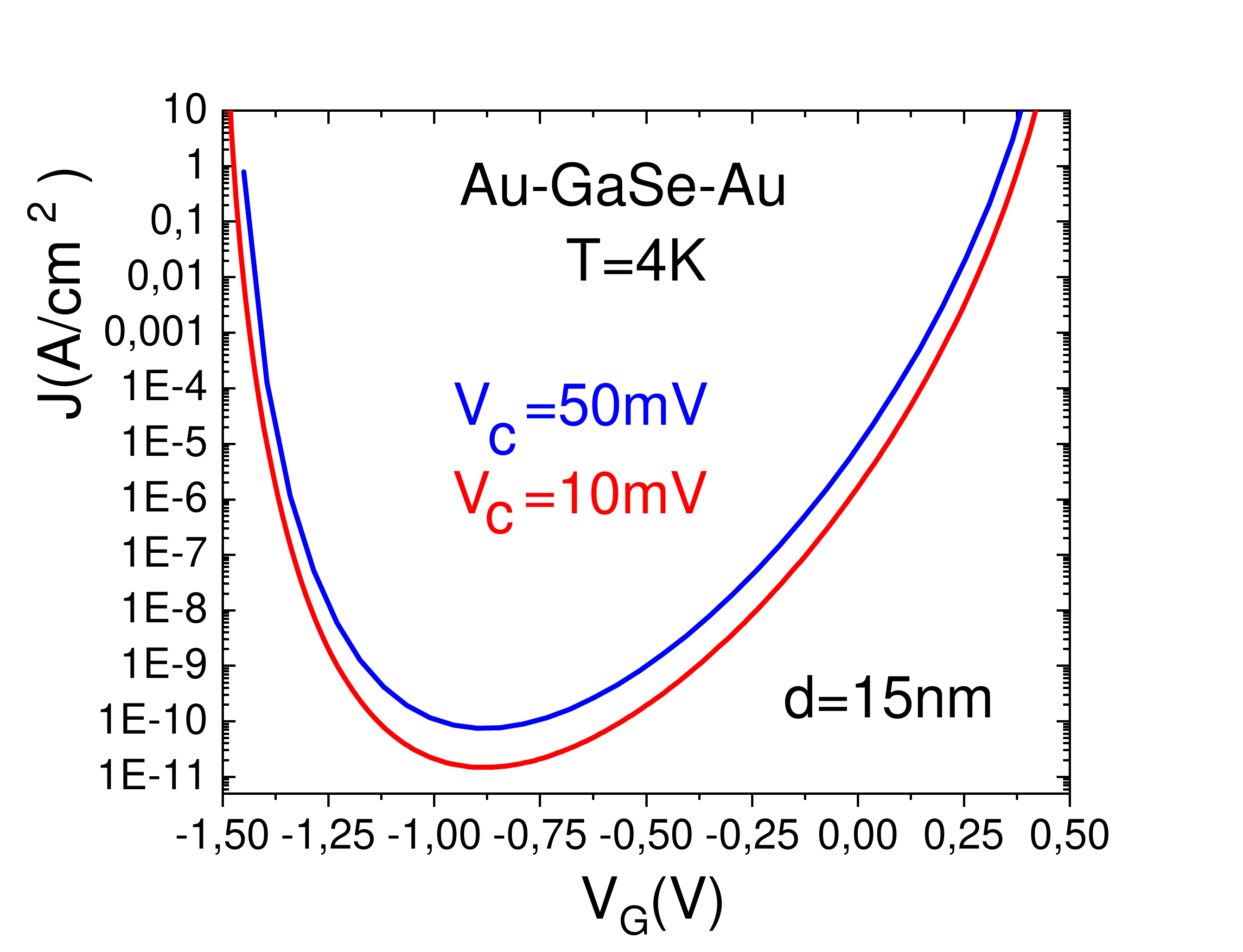}
\caption{The theoretical tunneling current $J$ versus gate voltage $V_G$ applied to the GaSe barrier. The curves are calculated for values of voltage $V_C$ = 50 meV and 10 meV applied to Au-GaSe-Au structure. The width of the GaSe barrier $d$ = 15 nm and $V^L_b$ = $V^R_b$=1.48 eV are used. The asymmetrical shape of the curves is the result of the asymmetry energy dispersion $E(k_z^2)$ in the forbidden gap of GaSe, see Fig. 3.} \label{fig7th}
\end{figure}
\section{\label{sec:level1} Principles of operation of the proposed tunneling transistor\protect\\ \lowercase{}}

The basis of the proposed transistor is the observation that the current that flows through the Source-Channel-Drain structure biased with the constant voltage $V_C$ can be changed depending on the magnitude of the gate voltage $V_G$ applied to the channel by an electrode separated from the channel by the oxide layer, see e. g. Ref. [10].
In other words, by increasing or lowering the potential energy of the channel in relation to the source, we can control $k_z^2$ of the current electrons, and thus the magnitude of the tunneling current, see Figs. 2-4 and Fig. 5. So, the modified formula for the $|k_z(z)|^2$ of an electron in the forbidden channel gap with the applied voltage $V_G$ looks like this
\begin{equation}
|k_z(z)|^2 = \left(1-\frac{E(z)-eV_G}{E_g}\right)(E(z)-eV_G)\frac{2m^*_0}{\hbar^2} + k^2_{\perp}\;.
\end{equation}
\begin{figure}
	\includegraphics[scale=0.4,angle=0, bb = 650 10 60 560]{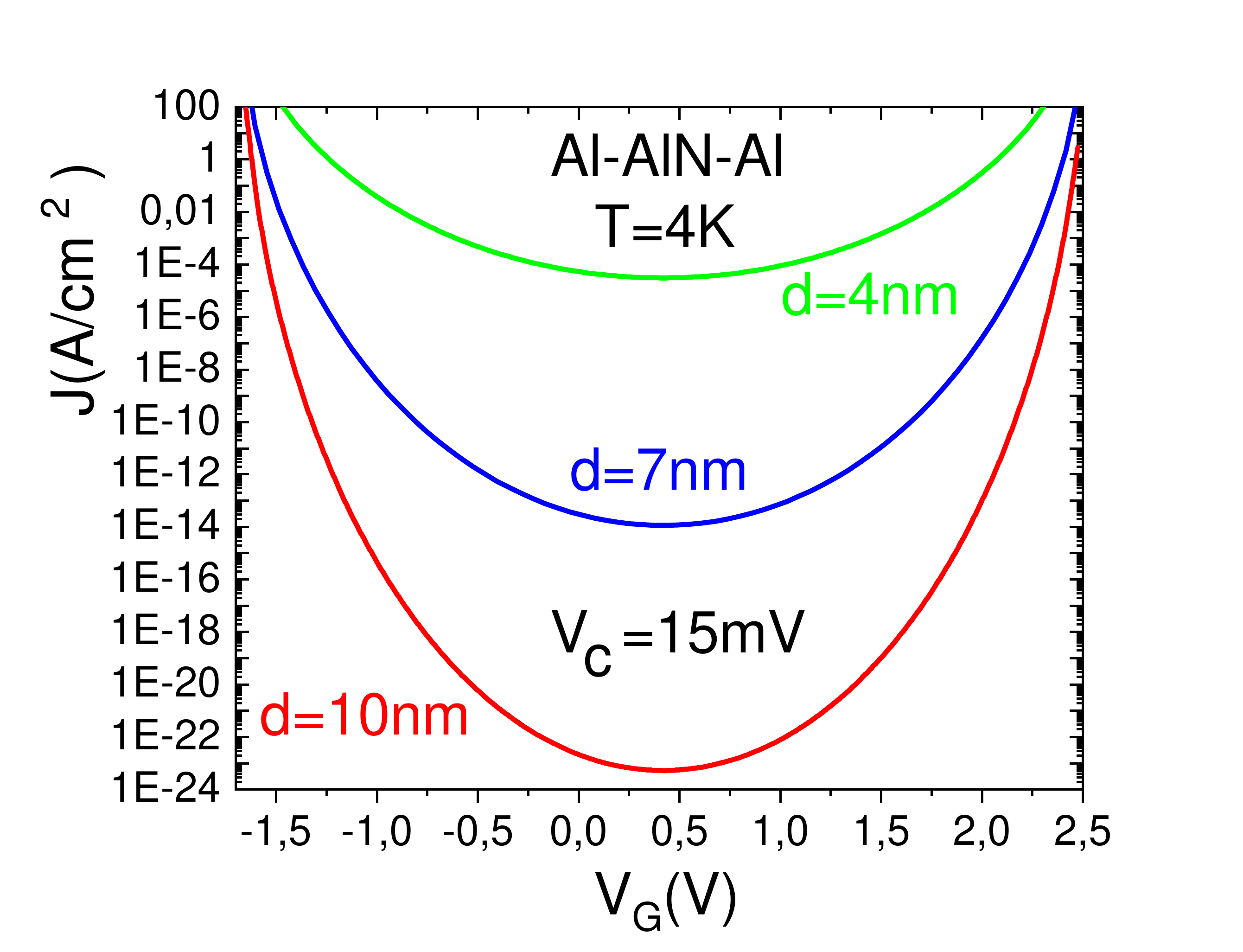}
	\caption{The theoretical tunneling current $J$ versus gate voltage $V_G$ applied to the AlN barrier. The curves are calculated for value of voltage $V_C$ = 15 meV applied to the  Al-AlN-Al structure. The widths of the barrier $d$ = 4 nm, 7 nm and 10 nm and $V^L_b$ = $V^R_b$=1.68 eV are used.} \label{fig8th}
\end{figure}

The dependences $J(V_G)$ for the InAs-InAs-InAs, Au-GaSe-Au and Al-AlN-Al structures calculated using Eqs. 7 and 9 i.e. at helium temperature are presented in Figs. 6-8. A negative or positive value of $V_G$ applied to the GaSe element in the Au-GaSe-Au structure or  to the AlN element in the  Al-AlN-Al structure means a reduction or increase of the virtual shift of $V_b^L$ and $V_b^R$ of the GaSe barrier with respect to the Au source or the AlN barrier with respect to the Al source and thus the change $k_z^2$ of current electrons.  The obvious relationship that the wider the channel, the smaller the minimum tunnel current $J$ is shown in Fig. 8. Comparing the relationship $J(V_G)$ with $E(k_z^2)$ (Figs. 2-4) for these structures we see that the wider the channel bandgap is, the larger range of $k_z$ values it has, and thus a greater range of tunnel current changes is possible.
\begin{figure}
	\includegraphics[scale=0.35,angle=0, bb = 650 10 60 560]{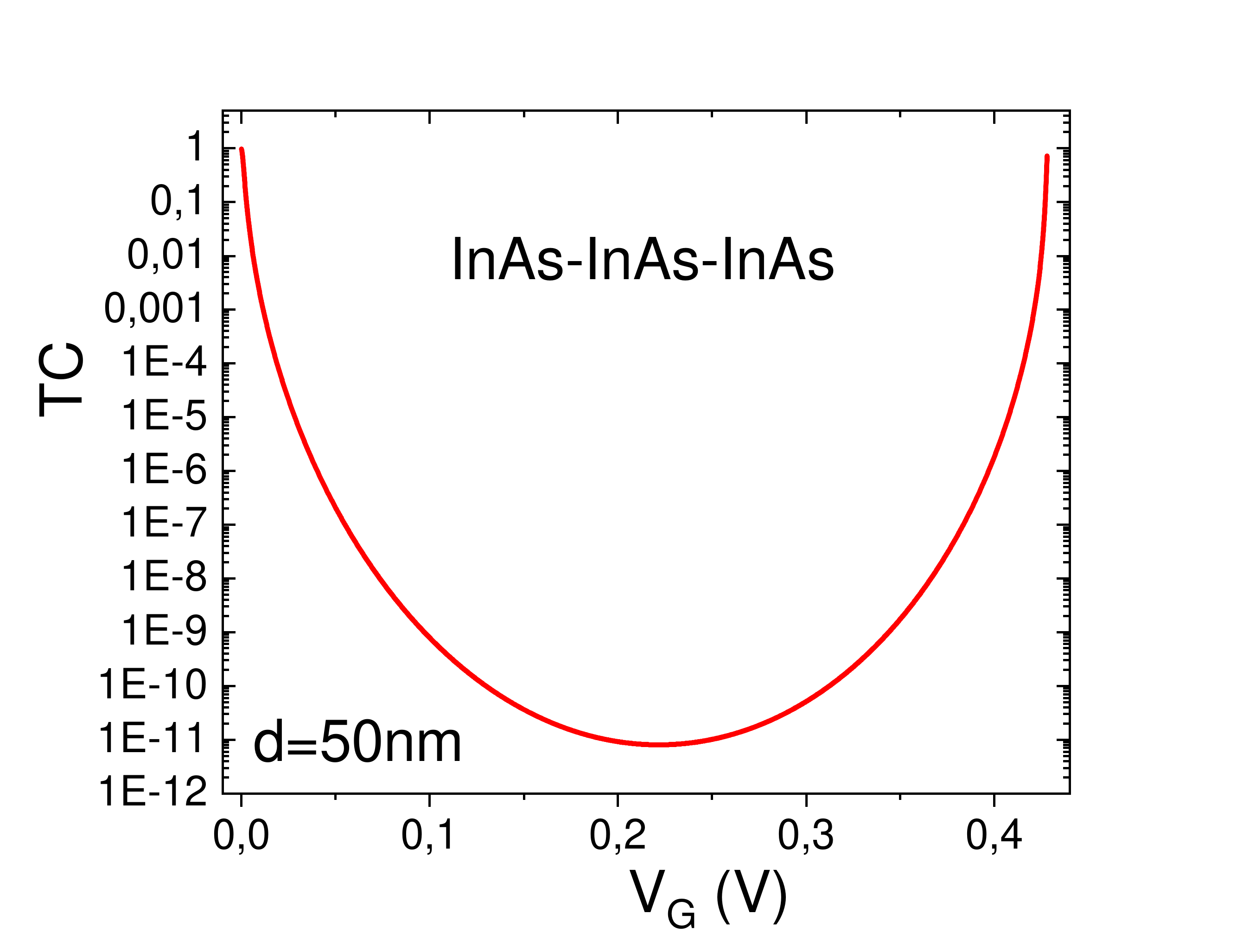}
	\caption{Transmission probability $T_C$ for electrons with a specific value of $E$ within forbidden gap of InAs versus barrier potential $V_G$. The curve is calculated for values of $V_G$ in the range 0 - $E_g/e$.} \label{fig9th}
\end{figure}

The conclusion that can be drawn from the $J(V_G)$ curves in Figures 6-8 is as follows: the smaller $V_C$, the greater the ratio of the maximum tunnel current $J_{MAX}$ to the minimum tunnel current $J_{MIN}$. The reason is that the smaller $V_C$ and, consequently, $J_C$, the fewer electrons form the current and the smaller the difference between $|k_z|^2$ of these electrons is. Thus, the highest ratio of $J_{MAX}$ to $J_{MIN}$ for a given width of $d$ will occur when the voltage $V_C$ will be extremely small, i.e. when the tunneling current will be formed exclusively from electrons of the same energy $E$. In this case, it is convenient to calculate the transmission coefficient, $TC$, of electrons tunneling through the forbidden gap of the channel vs. $V_G$, (the procedure is included e.g. in Ref.\cite{Pf}). Such a dependence $TC(V_G)$ for the InAs-InAs-InAs structure is shown in Fig. 9. It can be seen that the ratio of $TC_{MAX}$ to $TC_{MIN}$ and therefore $J_{MAX}$ to $J_{MIN}$ is indeed extremely large.
\begin{figure}
	\includegraphics[scale=0.35,angle=0, bb = 650 10 60 560]{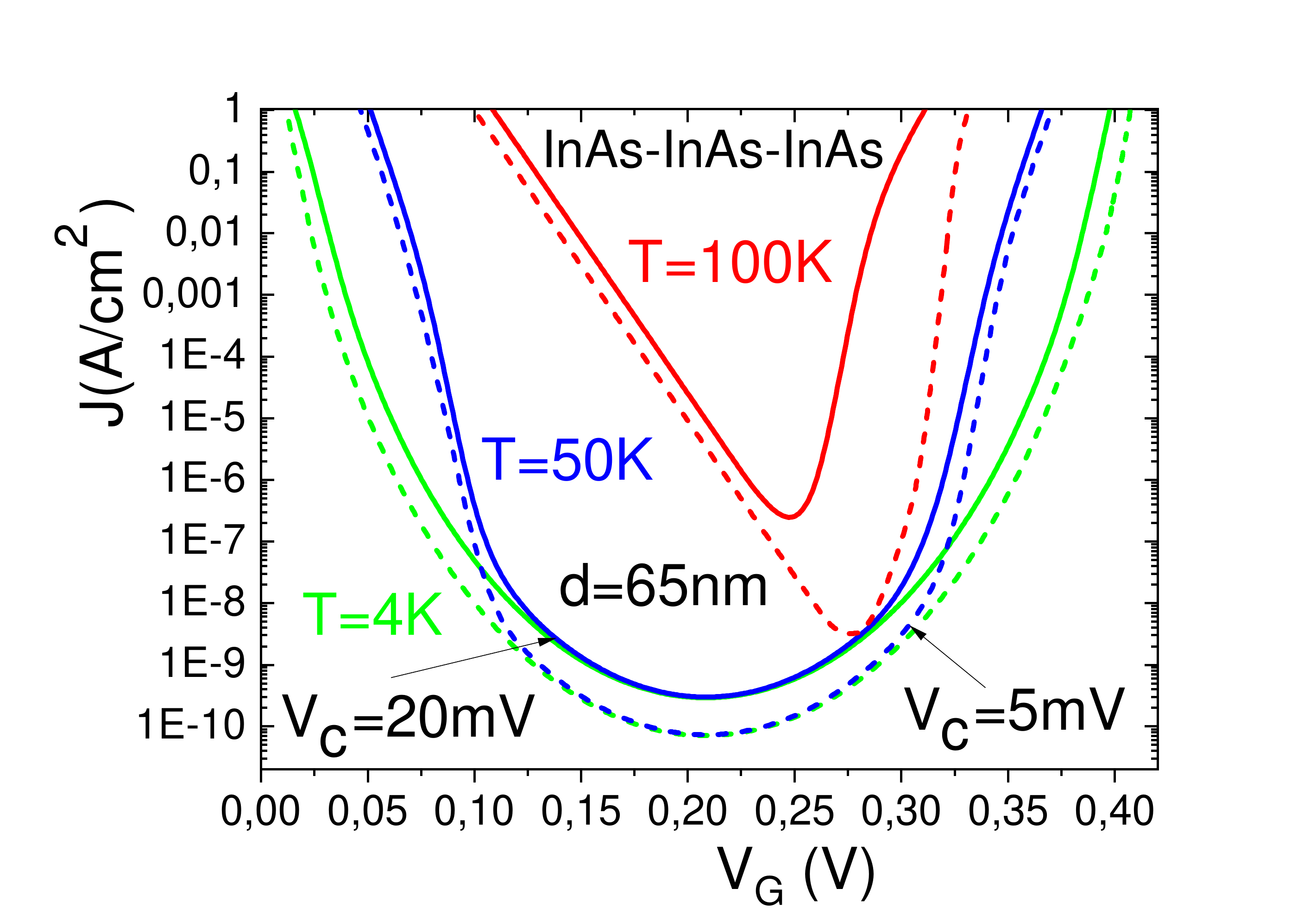}
	\caption{The theoretical tunneling current, $J$ versus gate voltage, $V_G$, calculated for different values of temperature for the InAs-InAs-InAs structure, i.e. with narrow gap channel, for applied voltage $V_C$ = 20 mV (solid lines) and for $V_C$ = 5 mV (dashed lines). The InAs channel forbidden gap $E_g$ = 0.417 eV and its width $d$ = 15 nm are used. It is seen that the change in $J(V_G)$ characteristic with increasing temperature above T = 50 K becomes more and more significant.} \label{fig10th}
\end{figure}

\section{\label{sec:level1} Dependence of the $J(V_G)$ characteristic on temperature \protect\\ \lowercase{}}
\begin{figure}
	\includegraphics[scale=0.35,angle=0, bb = 650 10 60 560]{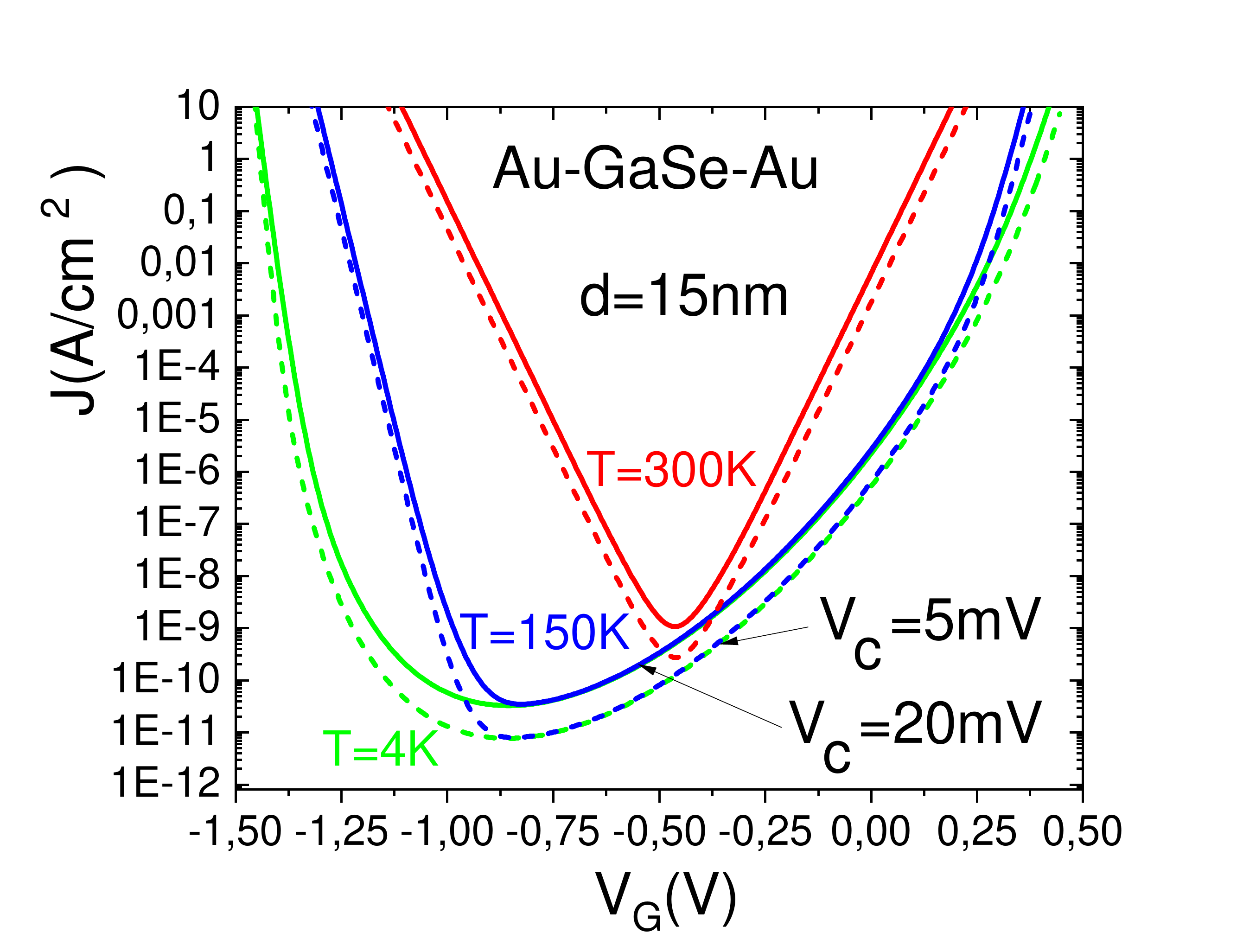}
	\caption{The theoretical tunneling current, $J$, versus gate voltage, $V_G$, calculated for different values of temperature for the structure  Au-GaSe-Au, i.e. with medium gap channel, for applied voltage $V_C$ = 20 mV (solid lines) and for $V_C$ = 5 mV (dashed lines). The GaSe channel forbidden gap $E_g$ = 2 eV and its width $d$ = 15 nm are used. It is seen that the change in $J(V_G)$ characteristic with increasing temperature  above T = 200 K becomes significant.} \label{fig11th}
\end{figure}

With increasing temperature, according to the Fermi-Dirac distribution function $f_{F-D}$, the number of electrons with the energy greater than $E_F$ increases. This means that the increasing number of electrons passes the channel of the studied structure not through its band gap but through the conduction band. This in turn means that the contribution of tunneling electrons to the total current, $J$, decreases the more the narrower the forbidden gap of the channel is. As a result, the characteristic $J(V_G)$ calculated using Eqs.(6) in the example structure with a narrow band gap channel changes significantly for the temperature above T = 50 K, see Fig. 10. In the structure with a medium band gap $J(V_G)$ changes significantly for the temperature above about T = 200 K, see Fig. 11. While in the structure with a wide band gap studied up to T = 300 K $J(V_G)$ changes slightly, see Fig. 12.
\section{\label{sec:level1} Summary\protect\\ \lowercase{}}
\begin{figure}
	\includegraphics[scale=0.35,angle=0, bb = 650 10 60 560]{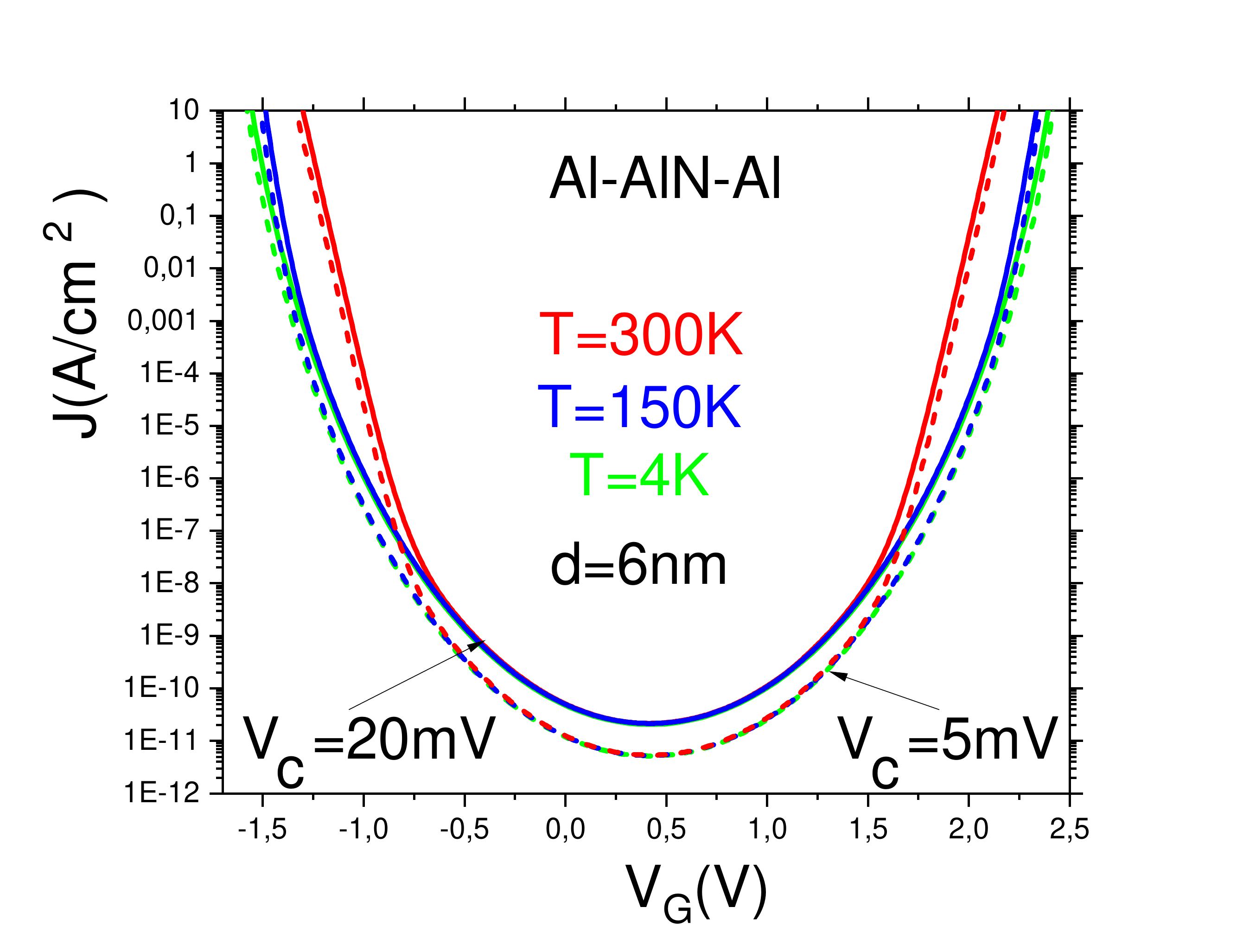}
	\caption{The theoretical tunneling current $J$ versus gate voltage $V_G$ calculated for different values of temperature for the structure Al-AlN-Al, i.e. with wide gap channel, for applied voltage $V_C$ = 20 mV  (solid lines) and for $V_C$ = 5 mV (dashed lines). The AlN channel forbidden gap $E_g$ = 4.2 eV and its width $d$ = 6 nm are used. It can be seen that the very strong dependence of $J$ on $V_G$ is preserved at least up to temperature of T = 300 K.} \label{fig12th}
\end{figure}

The presented theoretical project of controlled transmission of electron current through the barrier potential can be treated as a new type of tunnel transistor without a heterojunction unlike classic TFET. It is a simple extension of the experimentally proven structure for studying the dependence of tunnel current on applied voltage (e.g. Refs. 2-7). Its mechanism of operation (see Refs. 10-12) consists in controlling the height of the potential barrier (channel) polarized with voltage $V_G$, i.e. controlling the amount of current $J$ that tunnels through this barrier.
 The $J$ change for a channel with a specific band gap value, is bigger, the bigger is $m_C$ or $m_V$ or both see Figs. 3 and 7. The maximum change in $J$ formed by tunneling electrons along the entire length of the channel, computed at helium temperature, is due to a change in $V_G$ in the range 0 - ($E_g/2e-V_C)$. For example, for $V_C$ = 10 meV in the InAs structure ($d$ = 65 nm), a 180 mV change in $V_G$ means a change in $J$ by 9 orders of magnitude, for $V_C$ = 10 meV in the Au-GaSe-Au structure ($d$ = 15 nm), a change in $V_G$ by 0.6 V means a change in $J$ at least by 12 orders of magnitude, while for $V_C$ = 15 meV in the  Al-AlN-Al structure ($d$ = 10 nm), a change in $V_G$ by 2 V means a change in $J$ at least by 24 orders of magnitude. In addition, one can notice that dependence $J$ on $V_G$ can be adjusted by changing the voltage $V_C$ and width $d$ of the channel.

Changes in the $J(V_G)$ characteristic with increasing temperature depend on the width of the band gap of the channel and are the smaller, the wider band gap is. A channel band gap of the order of 3 eV or more means that the relationship $J(V_G)$ in the 4 K - 300 K range is almost unchanged, because even at the temperature T = 300 K there are practically no electrons with the energy about $E_F+E_G/2$, i.e. almost all electrons tunnel.

The basis of this project is the unchangeable dependence of the electron wave vector $k_z$ on its energy in the band gap of the channel in the considered structure. This means that a possible slight deviation of the actual shape of the barrier potential or the values of structure parameters from those adapted for calculations may cause minor quantitative changes in the relation $J(V_G)$, but not qualitative ones.

\ \\


\section*{References}
\begin{thebibliography}
{99}\label{sec:TeXbooks}
\bibitem{Fr1930} J. Frenkel, Phys. Rev. \textbf{36}, 1604 (1930).
\bibitem{Si} J. G. Simmons, J. Appl. Phys. \textbf{34}, 1793 (1963).	
\bibitem{St} R. Stratton, G. Lewicki and C. A. Mead, J. Phys. Chem. Solids.\textbf{ 27}, 1599 (1966).
\bibitem{Pa} G. H. Parker and C. A. Mead, Phys. Rev. Lett. \textbf{21}, 605 (1968).
\bibitem{Ku} S. L. Kurtin, T. C. McGill, and C. A. Mead, Phys. Rev. B \textbf{3}, 3368 (1971).
\bibitem{Gill}  T. C. McGill, PhD thesis, California Institute of Technology Pasadena California (1969).
\bibitem{Zh} P. Zhang, Scientific Reports\textbf{ 5}, 9826 (2015).
\bibitem{Kl} O. Klein, Z. Phys.\textbf{ 53}, 157 (1929).
\bibitem{Ca} A. Calogeracos and N. Dombey, Contemp. Phys. \textbf{40}, 313 (1999).
\bibitem{Ka} M. I. Katsnelson, K. S. Novoselov and A. K. Geim, Nat. Phys.\textbf{ 2}, 620 (2006).
\bibitem{Li2017} Zhenglu Li, Ting Cao, Meng Wu, and Steven G. Louie, Nano Letters \textbf{17}, 2280 (2017).
\bibitem{Pf} P. Pfeffer, W. Zawadzki and K. Dybko, Semicond. Sci. Technol. \textbf{36}, 045023 (2021).
\bibitem{Sa} D. Sarkar et al, Nature \textbf{526}, 91 (2015).
\bibitem{Av} U. E. Avci, D. H. Morris, and I. A. Young, J. Electron Devices Soc. \textbf{3}, 88 (2015).
\bibitem{Ve} D. Verreck, A. S. Verhulst, and G. Groeseneken, \textit{The Tunnel Field-Effect
Transistor}, Wiley Encyclopedia of Electrical and Electronics
Engineering, 1-24, (2016), library.wiley.com/doi/10.1002/047134608X.W8333.
\bibitem{Il} H. Ilatikhameneh, G. Klimeck, J. Appenzeller, and R. Rahman, arXiv:1603.09402v1 [cond-mat.mes-hall] 2016.
\bibitem{Co} C. Convertino et al, J. Phys. Condens. Matter \textbf{30}, 264005 (2018).
\bibitem{Fr1952} W. Franz, (in German), Ann. Phys.\textbf{ 6}, 17 (1952).
\bibitem{Ta} Y. Taur and J. Wu, IEEE Transactions on Electron Devices, \textbf{ 63}, 869 (2016).
\end{thebibliography}
\end{document}